# Transverse magneto-gyrotropy driven by spin-induced ferroaxiality

Y. Okamura[1,2†,*], T. Kagawa[1†], A. Nakamura[3], S. Nishihaya[3], M. Uchida[3,4], S. Minami[5,6], R. Arita[5,7] and Y. Takahashi[1*]

[1]*Department of Applied Physics and Quantum Phase Electronics Centre, University of Tokyo, Tokyo 113-8656, Japan*

[2]*Research Center for Advanced Science and Technology, University of Tokyo, Tokyo 153-0041, Japan*

[3]*Department of Physics, Institute of Science Tokyo, Tokyo 152-8550, Japan*

[4]*Toyota Physical and Chemical Research Institute, Nagakute 480-1192, Japan*

[5]*RIKEN Centre for Emergent Matter Science (CEMS), Wako 351-0198, Japan*

[6]*Department of Mechanical Engineering and Science, Kyoto University, Kyoto, Japan*

[7]*Department of Physics, University of Tokyo, Bunkyo-ku, Tokyo, Japan*

[†]These authors equally contributed to this work

[*]To whom correspondence should be addressed (okamura@ap.t.u-tokyo.ac.jp, youtarou-takahashi@ap.t.u-tokyo.ac.jp)

**Photon chirality is inherently sensitive to structural handedness and magnetism, which break spatial inversion and time-reversal symmetries, respectively[1-6]. Ferroaxial order, by contrast, hosts two-dimensional chirality while preserving both symmetries, thereby precluding its direct coupling to photon chirality[7-14]. Intertwining ferroaxiality with magnetic order, however, may lift such constraints and enable unconventional coupling to photon chirality[15-18]. Nevertheless, the optical response arising from such intertwined order has yet to be demonstrated experimentally. Here, we demonstrate transverse magneto-gyrotropy arising from spin-driven ferroaxiality in the Weyl magnet $EuCd_2Sb_2$. Whereas the conventional polar magneto-optical Kerr effect is governed by magnetization parallel to the light-propagation direction, we observe a gyrotropic response associated with magnetization transverse to the light-propagation direction. The optical rotation exhibits a characteristic 120° periodicity upon rotation of the in-plane magnetic field, reflecting the threefold symmetry of the emergent ferroaxial state and revealing the ferroaxial origin of the transverse gyrotropic response. This coupling to photon chirality enables vectorial control of the gyrotropic response through the magnetic field direction. Our findings establish an interaction between photon chirality and transverse magnetization enabled by intertwined ferroaxial and magnetic orders, demonstrating a novel form of multiferroicity with functional light-matter interaction.**

Light carries not only energy and momentum but also chirality encoded in its polarization. In particular, circularly polarized photons possess a well-defined helicity, representing the handedness of their spin angular momentum projected onto the propagation direction. The interaction of this photonic chirality with matter provides a powerful means of probing and controlling electronic, magnetic, and structural handedness as well as time-reversal symmetry breaking[1-6]. Its manifestations range from natural optical activity and circular dichroism in chiral media to magneto-optical effects in systems with broken time-reversal symmetry. These phenomena arise when matter distinguishes between photons of opposite helicities, thereby converting microscopic symmetry breaking into a measurable optical contrast.

In magnetic materials, helicity-dependent optical responses are conventionally associated with magnetic field or magnetization parallel to the light-propagation direction[3,5]. This relationship underlies magnetic circular dichroism and the polar magneto-optical Kerr effect (MOKE), both of which provide sensitive probes of time-reversal-symmetry breaking. More generally, however, the coupling between photon helicity and matter is governed not only by magnetic-field direction but also by the spatial symmetries of the material. In particular, imparting ferroaxiality to a magnetic system introduces an additional coupling channel beyond that associated solely with the longitudinal magnetization parallel to the light-propagation direction. Ferroaxiality represents a two-dimensional rotational arrangement of electric dipoles $\boldsymbol{P}_i$ located at positions $\boldsymbol{R}_i$ (Fig. 1a), which has recently attracted growing interest as a novel ferroic state distinct from ferroelectrics and ferromagnets[7-14]. The ferroaxial moment is characterized by an electric toroidal dipole, $\boldsymbol{G} \sim \sum_i \boldsymbol{R}_i \times \boldsymbol{P}_i$ (Fig. 1a), which is invariant under both time reversal and spatial inversion but breaks mirror symmetries containing its axis. This two-dimensional chirality acts as a rotator that converts an applied stimulus into a response perpendicular to it (Fig. 1a), which gives rise to a variety of off-diagonal physical responses[5,15,16]; an input field perpendicular to $\boldsymbol{G}$ direction can induce a conjugate response perpendicular to both. Accordingly, an applied magnetic field $\boldsymbol{B}$ can generate an effective field $\boldsymbol{B}_{\text{eff}}$ along $\boldsymbol{G} \times \boldsymbol{B}$, which induces the coupling to the chirality of the photon (Fig. 1b). It therefore establishes an unconventional coupling between photon chirality and transverse magnetization oriented perpendicular to the light-propagation vector, which leads to the time-reversal-odd optical rotation that is governed by the ferroaxial moment. Hereafter, we refer to this phenomenon as the transverse magneto-gyrotropy (TMG; see Fig. 1b), the transverse-magnetization counterpart of the conventional polar MOKE driven by magnetization parallel to the light-propagation vector. This novel optical effect, however, has yet to be established.

Despite these intriguing prospects, structurally driven ferroaxial materials remain rare and are often paramagnetic, limiting the magnitude of their magnetic responses. Moreover, unlike ferroelectric and ferromagnetic orders, ferroaxial order lacks an experimentally accessible conjugate field, posing a major obstacle to domain control and the realization of macroscopic ferroaxial responses. In this context, magnetically induced ferroaxiality, recently proposed theoretically, offers a promising route to overcoming these limitations[17,18]. In such systems, magnetic ordering lowers the non-ferroaxial symmetry of the underlying crystal and thereby induces ferroaxiality. This intertwining of ferroaxial and magnetic orders may not only enable magnetic control of the ferroaxial state but also enhance the associated magneto-optical responses.

Here we observe the TMG due to spin-driven ferroaxiality in the Weyl magnet $EuCd_2Sb_2$ using optical polarimetry[19-21]. Although its crystallographic structure is intrinsically non-ferroaxial, a transverse magnetic field induces an effective ferroaxial moment, which produces time-reversal-odd optical rotation. The coupling of this field-induced ferroaxial moment to the underlying threefold crystalline symmetry enables rotational control of the TMG. This contrasts with conventional MOKE, which is governed by the longitudinal magnetization component parallel to the light-propagation direction (Fig. 1c).

$EuCd_2Sb_2$ crystallizes in a trigonal structure with space group $P\bar{3}m1$, which does not belong to a ferroaxial symmetry class[19-21]. The magnetic Eu layers and itinerant $Cd_2Sb_2$ blocks are alternately stacked along the out-of-plane [001] direction, as shown in Fig. 1d. Below $T_N \sim 7.4$ K, $EuCd_2Sb_2$ exhibits an A-type antiferromagnetic order at zero magnetic field, in which the Eu spins are ferromagnetically aligned within the triangular-lattice plane and antiferromagnetically stacked along the [001] direction (see also the phase diagram in Supplementary Fig. 1). The crystal structure possesses $C_3$ rotational symmetry about the [001] direction, $C_2$ rotational symmetry along the in-plane [100] direction and its symmetry-equivalent directions, as well as mirror planes perpendicular to these axes. Notably, when viewed along the [100] direction, the Cd and Sb atoms form a ferroaxial-like arrangement characterized by a twofold rotation axis along [100] and a mirror plane perpendicular to this axis (Fig. 1d). Therefore, when the magnetic field is applied perpendicular to the [100] direction, the induced magnetization breaks the remaining twofold axes and mirror planes and the magnetic point group is reduced to 2'/$m$' from the original crystalline symmetry (Fig. 1e), corresponding to a ferroaxial state with broken time-reversal symmetry.

Importantly, this emergent ferroaxial state can be controlled by magnetic-field orientation. When the field is applied perpendicular to the other symmetry-equivalent ⟨100⟩ directions, the ferroaxial moment is correspondingly aligned perpendicular to the magnetic-field direction (Fig. 1e; see also Supplementary Fig. 2). As a consequence, the effective field $\boldsymbol{B}_{\mathrm{eff}}$ ($\parallel \boldsymbol{G} \times \boldsymbol{B}$) emerges along the [001] direction and reverses sign upon field reversal, giving rise to a 120° periodicity of $\boldsymbol{B}_{\mathrm{eff}}$ upon rotating the magnetic field $\boldsymbol{B}$ about the [001] axis, as summarized in Fig. 1f. This $\boldsymbol{B}_{\mathrm{eff}}$ induces the TMG with out-of-plane light incidence under an in-plane magnetic field (Fig. 1b), in sharp contrast to the conventional MOKE (Fig. 1c).

To demonstrate this, we experimentally measured the optical rotation spectra in the visible region under a rotating in-plane magnetic field of 0.34 T (Fig. 2a; see also Methods). Here, we define the $x$ and $z$ axes along the [100] and [001] directions, respectively, and $\varphi$ denotes the in-plane magnetic field direction measured from the [100] axis (see also Fig. 1f). The optical rotation angle, $\theta_{\mathrm{K}}$, was obtained by antisymmetrizing the polarization rotation angles measured under positive and negative magnetic fields. Clear resonant features are observed in the $\theta_{\mathrm{K}}$ spectra when the magnetic field is applied along the $y$ direction ($\varphi = 90°$; blue curve in Fig. 2b). This resonance is highly sensitive to the magnetic-field direction. Similar structure is observed for $\varphi = 150°$, but with a completely reversed sign (red curve in Fig. 2b). In contrast, $\theta_{\mathrm{K}}$ is nearly zero for $\varphi = 0°$ (gray curve in Fig. 2b). These observations are well consistent with the symmetry consideration shown in Fig. 1f.

We further perform continuous $\varphi$-dependent measurements at a fixed photon energy of 2.2 eV (Fig. 2c). The $\theta_{\mathrm{K}}$ signal exhibits a clear 120° periodicity reflecting the $C_3$ rotational symmetry about the $z$ axis, and its phase is locked to the crystal orientation. The amplitude of the periodic signal decreases with increasing temperature and abruptly disappears above $T_N$ (Figs. 2c–f). These observations demonstrate the emergence of a TMG response governed by an emergent ferroaxial moment intertwined with the transverse magnetization.

We note that the optical rotation is also observed when the incident light polarization is rotated by 90°, namely for $E^{\omega} \perp x$. The obtained $\theta_{\mathrm{K}}$ spectrum for $E^{\omega} \perp x$ quantitatively coincides with that for $E^{\omega} \parallel x$ (red and orange curves in Fig. 3a). This agreement demonstrates that the observed rotation originates from the antisymmetric off-diagonal components of the dielectric tensor, in accord with the gyrotropic response, while indicating that polarization-dependent linear optical anisotropy, such as the Cotton–Mouton effect, does not make a substantial contribution[22,23].

It should be emphasized that the TMG induced by the in-plane magnetic field continues to increase even above the saturation field of the longitudinal magnetization parallel to the applied field (Fig. 3b). This distinct field dependence suggests that the TMG essentially probes an out-of-plane magnetization induced by an effective field $\boldsymbol{B}_{\mathrm{eff}}$ ($\parallel \boldsymbol{G} \times \boldsymbol{B}$), rather than the in-plane magnetization along the applied field. This out-of-plane magnetization, which is perpendicular to the applied field, is likely predominantly orbital in origin because the spin moments are expected to align nearly parallel to the applied field owing to the weak magnetic anisotropy of $Eu^{2+}$ (Supplementary Fig. 3).

To gain further microscopic insights into the TMG, we measured the MOKE spectrum when the magnetic field is applied parallel to the light incidence (Fig. 1c and 3c). The magneto-optical rotation spectra exhibit distinct resonance energies and spectral features for the out-of-plane and in-plane field configurations, with the out-of-plane field producing a substantially larger optical rotation than the in-plane field. Thus, assuming an isotropic magnetic susceptibility (Supplementary Fig. 3), the gyrotropic response for an arbitrary three-dimensional magnetic-field direction can be reconstructed, as summarized in Fig. 3d. Whereas conventional systems exhibit positive and negative Kerr signals separated by the equatorial plane, the present system displays an additional 120° periodic modulation along the azimuthal direction, superimposed on the conventional 180° periodicity along the polar direction.

The distinct spectral characteristics observed under the two field configurations indicate different underlying electronic properties. Since the Kerr response reflects the quantum geometry of the electronic structure[24–26], we calculated the *z*-component of Berry curvature induced by in-plane and out-of-plane magnetizations (Fig. 4; see also Methods); note that the Berry curvature provides useful insight into the underlying quantum geometry, although it is not quantitatively equivalent to the finite-frequency magneto-optical response measured here. Figures 4a–c show the momentum-space distribution of the Berry curvature summed over the occupied states under an in-plane magnetic field. The Berry-curvature distribution for φ = 150° (Fig. 4c) reflects the ferroaxial symmetry and can be approximately understood as that for φ = 90° (Fig. 4b) rotated by 60° following the magnetization direction and accompanied by a sign reversal of the local Berry curvature. As a result, the net Berry curvature also changes sign (Supplementary Fig. 4). By contrast, for φ = 0° (Fig. 4a), the Berry curvature remains locally finite, while its momentum-space integral vanishes (Supplementary Fig. 4). These field induced rotational variations of Berry curvature are qualitatively consistent with the experimentally observed characteristics of the TMG induced by the transverse magnetization (Fig. 2b).

We further calculated the Berry-curvature distribution under an out-of-plane magnetic field, which is qualitatively distinct from that obtained for the in-plane field (Fig. 4d). The markedly different Berry-curvature distributions should lead to the distinct resonance energies and spectral intensities in the optical rotation spectra, as observed for the in-plane and out-of-plane field configurations (Fig. 3a and 3b). Thus, the anisotropic gyrotropic response in Fig. 3d reflects the vectorial evolution of the quantum geometry in a ferroaxial magnet.

The present magneto-optical measurement reveals the magneto-optical gyration induced by the transverse magnetization. While it is well established that the natural optical activity is a sensitive probe of the three-dimensional crystalline chirality, our findings demonstrate the unconventional coupling between the chirality of photon and ferroaxiality or two-dimensional chirality intertwined with the magnetization. We also establish the spin-induced ferroaxial order with high controllability and functionality, which enables engineering of quantum geometric properties through vectorial control of magnetization. This novel multiferroic state, coupled ferroaxial and ferromagnetic orders, may host unique coupled elementary excitations accompanied by coherent fluctuations of both orders, similar to the so-called electromagnon in magnetoelectric multiferroics[27-30], which is potentially useful for ultrafast control of coupled order parameters via intense light irradiation[31]. Our results open a new avenue for exploring unconventional quantum geometric phenomena unique to ferroaxial magnets[18] and further suggest a promising platform for future multiferroic devices.

**Acknowledgement**

This work was partially supported by JSPS KAKENHI (24K00758, 25H01252, 25KK22214, 25H00841, 26H00594, 26H00646, and 26H01284), JST FOREST Program (grant nos. JPMJFR212X and JPMJFR202N), Toyota Riken Rising Fellow Program by

Toyota Physical and Chemical Research Institute, Japan, and STAR Award funded by the Tokyo Tech Fund, Japan.

**Author contributions**

Y.O. and Y.T. conceived the project. Y.O. and T.K. performed the optical measurement and analysed the data under supervision of Y.T. A.N. and S.N. grew the thin film under supervision of M.U. S.M. performed the first-principles calculation under supervision of R.A. All the authors discussed and interpreted the results. Y.O., T.K. and Y.T. wrote the manuscript with assistance of other authors.

**Additional information**

**Competing interests:** The authors declare no competing financial interests.

## Methods

**Thin film growth.** $EuCd_2Sb_2$ films were grown in an Epiquest RC1100 chamber on single-crystalline (111)A CdTe substrates. The in-plane lattice mismatch between $EuCd_2Sb_2$ and CdTe is 2.4 %. CdTe substrates were etched with 0.01% $Br_2$-methanol before loading them into the chamber and then heated to 750 ◦C with Cd flux supplied to obtain an atomically smooth surface. After annealing the substrate, it was cooled to the growth temperature of 360 ◦C. The beam equivalent pressures were measured by an ionization gauge and were set to $1.2 \times 10^{-5}$ Pa for Eu, $5.0\times 10^{-4}$ Pa for Cd, and $8.5 \times 10^{-6}$ Pa for Sb during the codeposition. This Cd-rich growth condition offers advantages in reducing carrier densities in $EuCd_2Sb_2$ films, because Cd deficiency is a major origin of the hole carriers. The films were then annealed in-situ at 500◦C for 5 min under exposure of Cd flux. The film thickness was set at 50 nm, and the growth rate was about 0.07 Å/s.

**Magneto-optical rotation measurement.** Optical rotation spectra were measured with use of a mononchromator-type spectrometer for 1.9 – 2.5 eV by balanced detection technique. To deduce the magneto-optical rotation spectra, we anti-symmetrized the spectra for the positive and negative magnetic fields. The in-plane magnetic field is applied using a Halbach magnet except for Fig. 3c; the high in-plane field is applied using a superconducting magnet. The out-of-plane magnetic field is applied using a room-temperature superconducting bore magnet.

**First-principles calculation.**

First-principles calculations of the band structure with spin-orbit coupling were performed using the VASP package[32-34] for experimentally determined film lattice parameters[19] (see also Supplementary Fig. 5). The generalized gradient approximation of PerdewBurke-Ernzerhof was adopted for the exchange-correlation functional[35]. The correlation effect was considered by a Hubbard U correction of U = 4.5 eV for the Eu-f orbitals[36]. A 16 × 16 × 10 k-point mesh with the Monkhorst-Pack scheme[37] was used for the Brillouin zone sampling of the primitive cell, and Gaussian smearing with a width of 0.02 eV was applied. From the Bloch states obtained in the first-principles calculations, a Wannier basis set was constructed using WANNIER90 and SymClosistWannier [38-41]. The basis consists of s, d, and f-character orbitals localized at the Eu site; s, p, and d-character orbitals at the Cd site; and s and p-character orbitals at the Sb site.

**Data availability.** The data that support the plots of this study are available from the corresponding author upon reasonable request.

## Figures

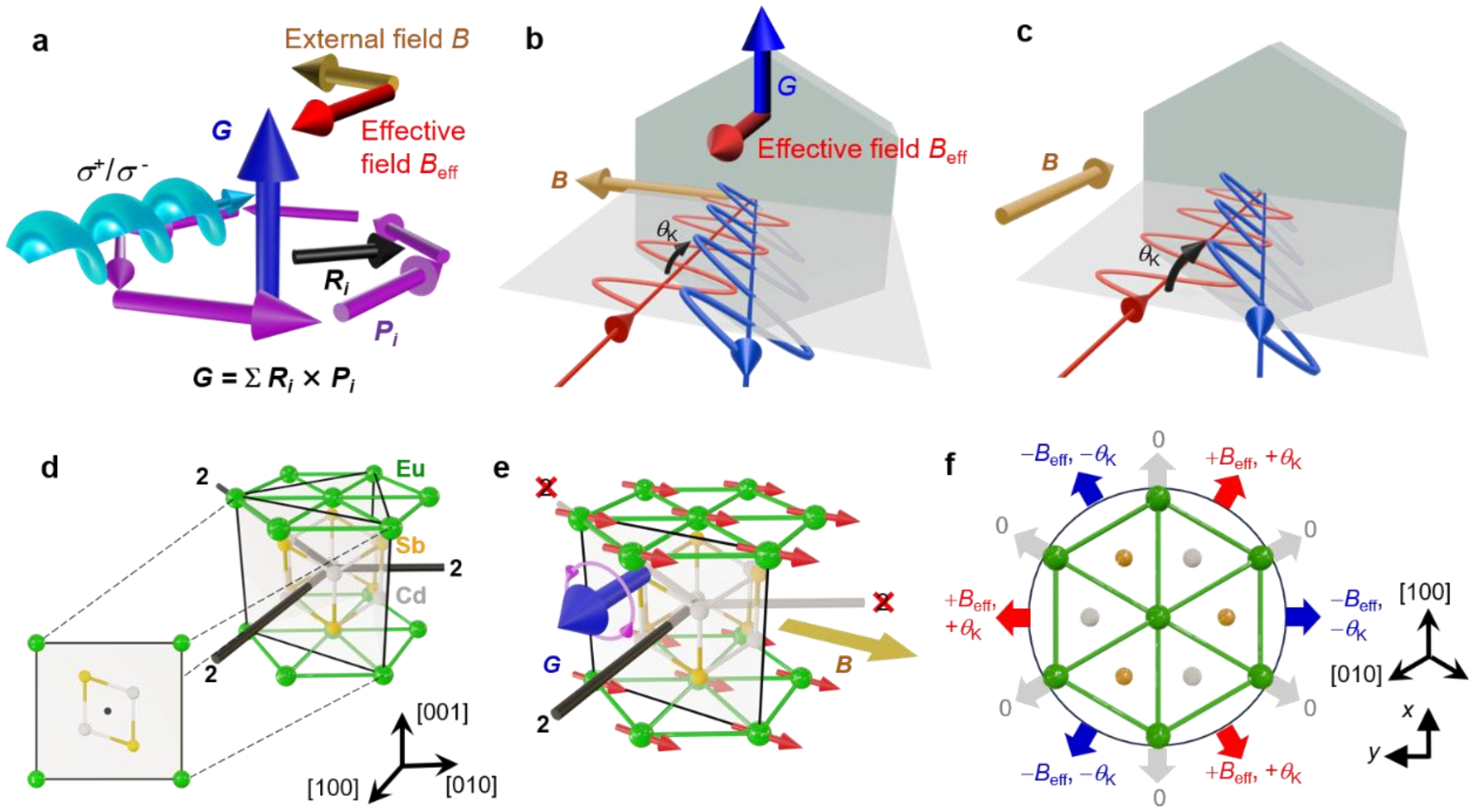

**Fig. 1| Spin-driven ferroaxial order exhibiting TMG. a,** Schematic illustration of electric toroidal moment $\boldsymbol{G}$. Rotational arrangement of electric dipoles $\boldsymbol{P}_i$ at positions $\boldsymbol{R}_i$ generates the $\boldsymbol{G}$. **b,** Schematic illustrations of the transverse magneto-gyration (TMG) under the transverse in-plane field $\boldsymbol{B}$. The effective magnetic field $\boldsymbol{B}_{\mathrm{eff}}$ and resultant TMG response $\theta_{\mathrm{K}}$ are also indicated. **c,** Schematic illustrations of conventional MOKE under the out-of-plane field. **d,** Crystal structure of $EuCd_2Sb_2$ with two-dimensional rotational arrangement of Cd and Sb atoms. The crystallographic axes are denoted in the hexagonal setting. **e,** Magnetization-induced generation of electric toroidal moment $\boldsymbol{G}$. $\boldsymbol{G}$ is characterized by twofold axis along [100] axis and mirror plane perpendicular to it, while other symmetries are broken by applying magnetic fields. **f,** Symmetry consideration of the effective magnetic field $\boldsymbol{B}_{\mathrm{eff}}$ and TMG response $\theta_{\mathrm{K}}$. The red, blue and gray arrows represent the field directions.

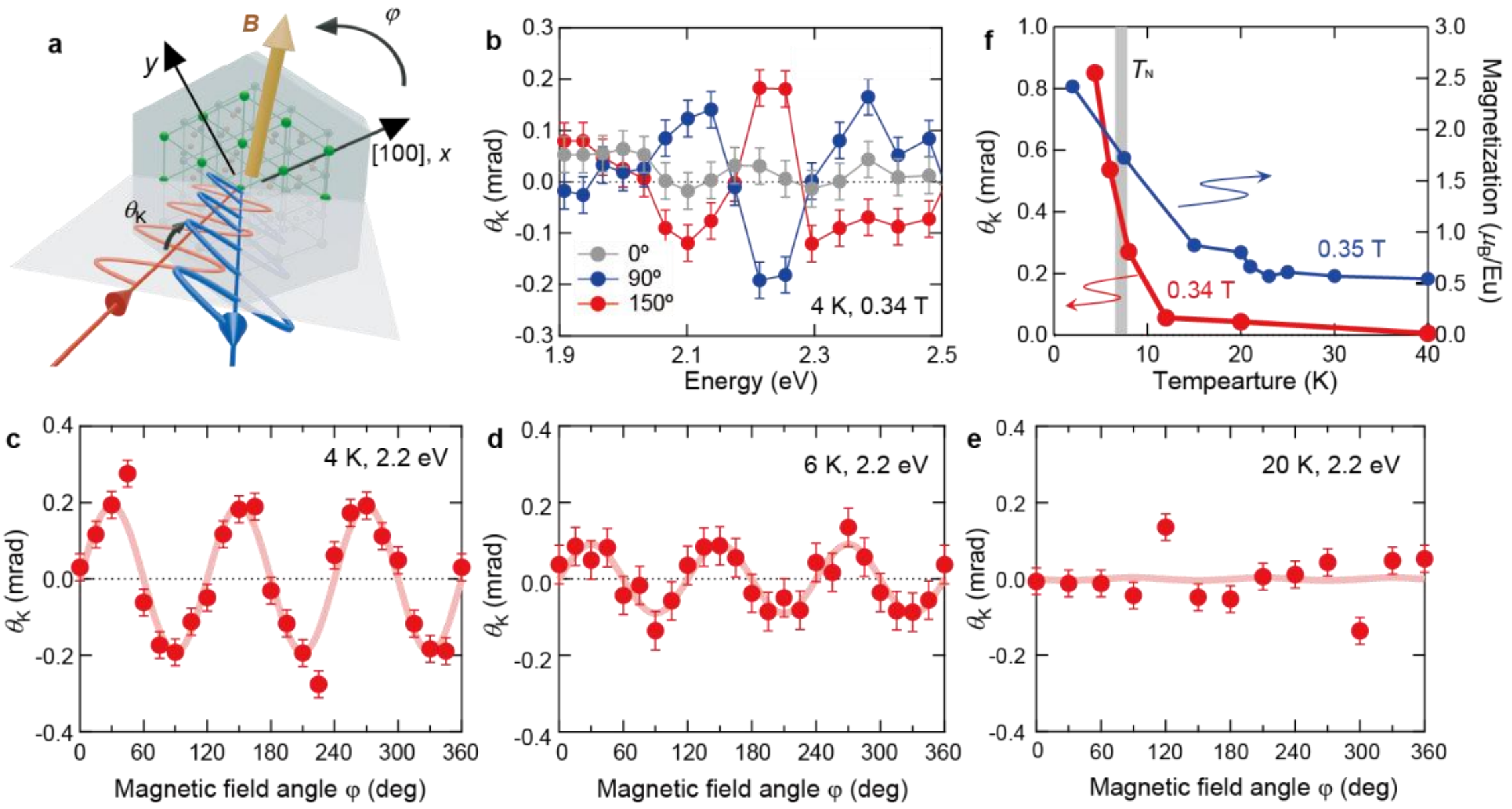

**Fig. 2| Observation of TMG. a,** Experimental setup of TMG induced by transverse magnetization. **b,** The optical rotation $\theta_K$ spectra for $\varphi = 0°$ (gray curve), 90° (blue curve), and 150° (red curve). **c-e,** Magnetic-field-angle dependence of $\theta_K$ at a fixed photon energy of 2.2 eV at each temperature (red symbols): 4 K (**c**), 6 K (**d**), and 20 K (**e**). The bold red curves represent the fits to $\theta_K$, showing the 120° periodicity. **f,** Temperature dependence of TMG (red symbols) and longitudinal magnetization along the magnetic field (blue symbols).

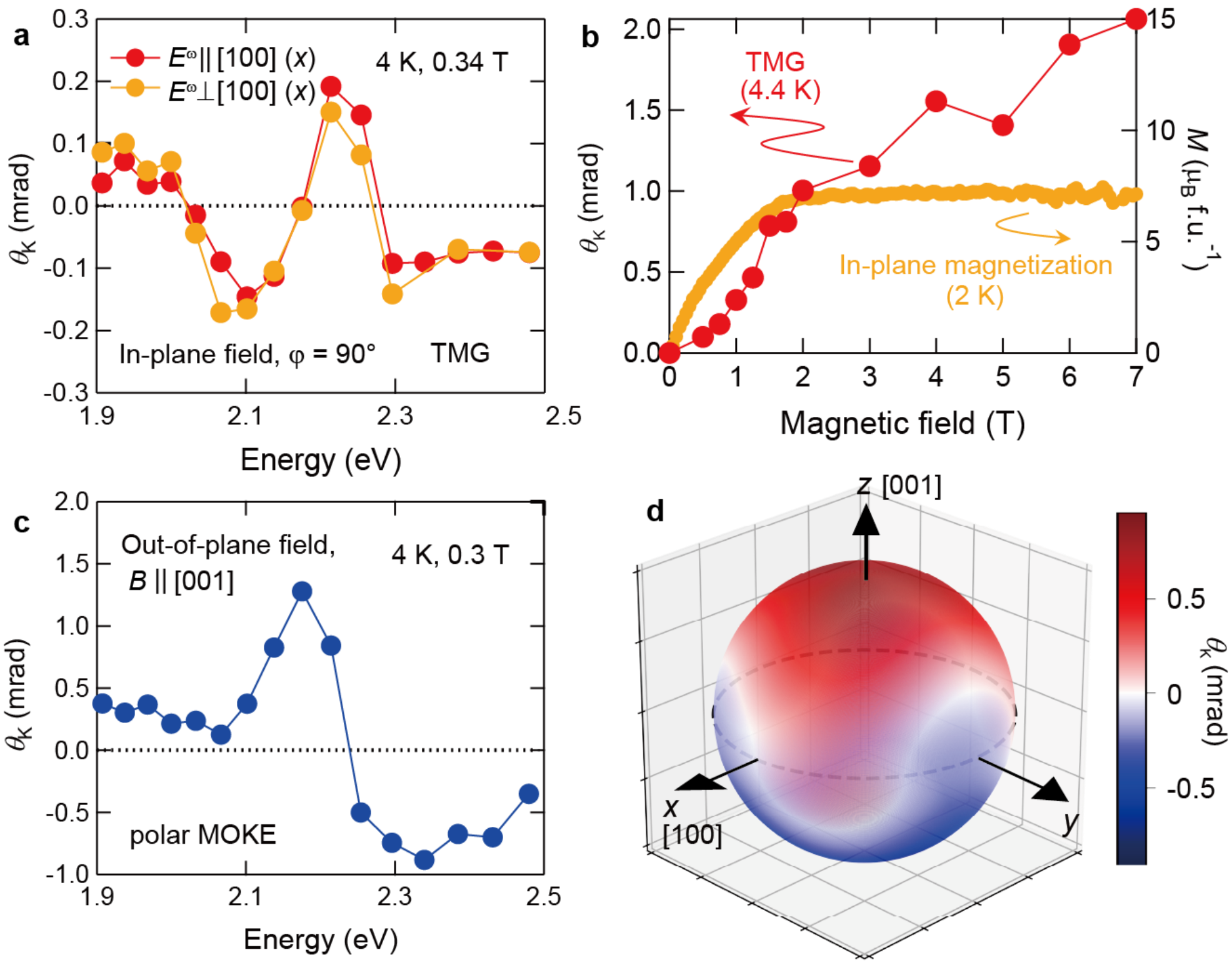

**Fig. 3| Anisotropic magneto-gyrotropy of ferroaxial magnet. a,** The TMG spectra due to transverse magnetization for two different incident polarization, $E^{\omega}\|[100]$ (red symbols) and $E^{\omega}\perp[100]$ (orange symbols). The measurement is performed in the configuration shown in Fig. 1b. **b,** Magnetic-field dependence of TMG (red symbols) and transverse magnetization (orange curve). **c,** The MOKE spectrum under the out-of-plane field. The measurement is performed in the configuration shown in Fig. 1c. **d,** Simulation of $\theta_K$ response at 2.2 eV and 0.34 T with respect to the three-dimensional field direction based on optical rotation data shown in a and c.

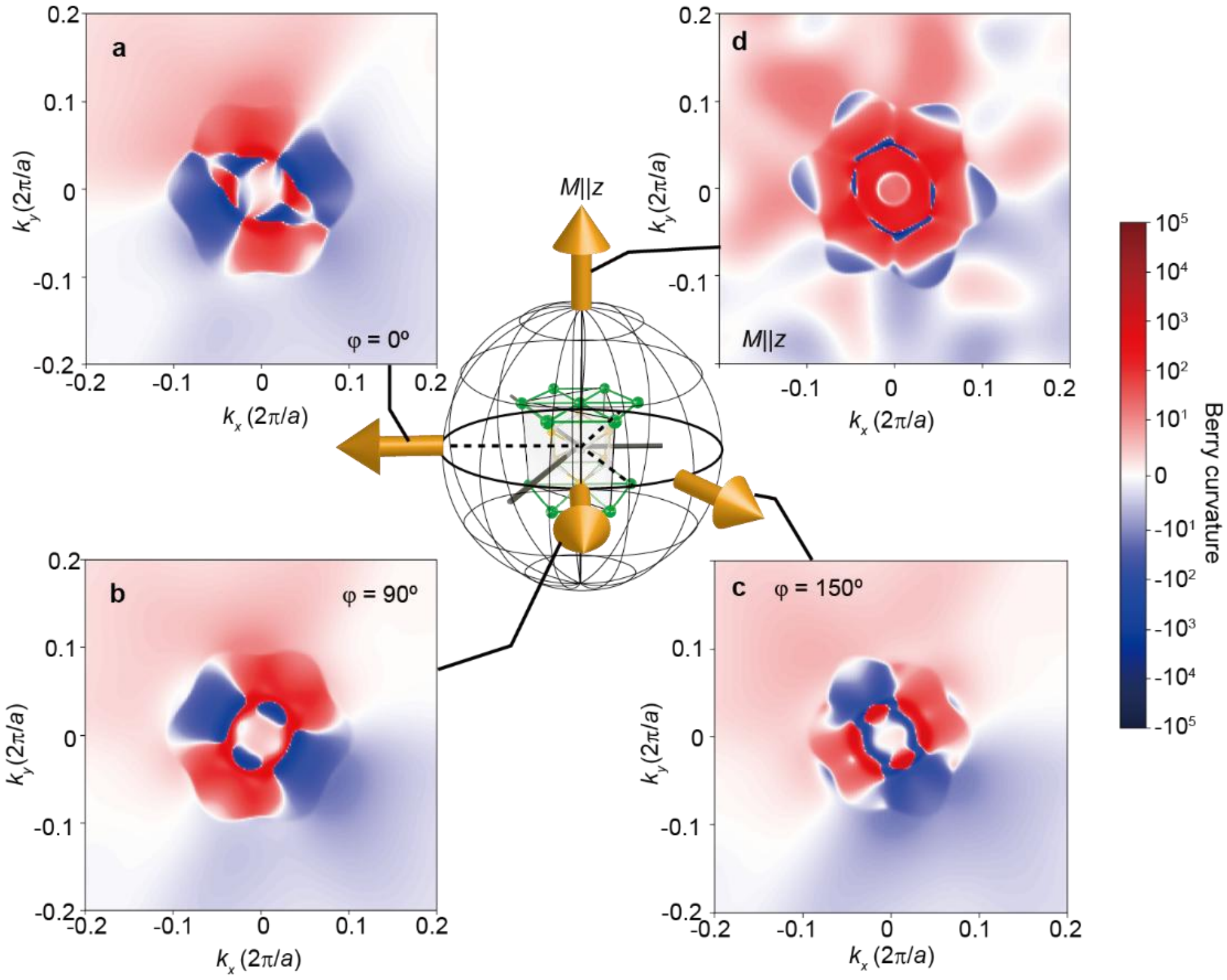

**Fig. 4| Vectorial control of quantum geometry in ferroaxial magnet. a-d,** The distribution of *z* component Berry curvature summed over the occupied states under in-plane and out-of-plane magnetic fields. Panels a, b and c show the cases for $\varphi = 0°$, 90°, and 150°. Panel d corresponds to the out-of-plane field case.